\documentclass[12pt]{article}
\usepackage{amsmath,amssymb,amsthm,amsxtra,overpic,bbm,bm,epsfig,ulem}
\usepackage{color}
\def\thefootnote{\fnsymbol{footnote}}

\begin{document}

\vspace{0.2cm}

\begin{center}
{\Large\bf The breaking of flavor democracy in the quark sector}
\end{center}

\vspace{0.2cm}

\begin{center}
{\bf Harald Fritzsch$^{a}$}, ~ {\bf Zhi-zhong Xing$^{b, c}$}, ~ {\bf
Di Zhang$^{b,d}$}\footnote{Email: dizhang@mails.ccnu.edu.cn} \\
{$^a$Physics Department, Ludwig Maximilians University, D-80333
Munich, Germany} \\
{$^b$Institute of High Energy Physics, and School of Physical
Sciences, \\ University of Chinese Academy of Sciences, Beijing 100049, China \\
$^c$Center for High Energy Physics, Peking University, Beijing
100080, China} \\
{$^d$College of Physical Science and Technology, Central China
Normal University, \\ Wuhan 430079, China}
\end{center}

\vspace{1.5cm}

\begin{abstract}
The democracy of quark flavors is a well-motivated flavor symmetry,
but it must be properly broken in order to explain the observed
quark mass spectrum and flavor mixing pattern. We reconstruct the
texture of flavor democracy breaking and evaluate its strength in a
novel way, by assuming a parallelism between the $Q=+2/3$ and
$Q=-1/3$ quark sectors and using a nontrivial parametrization of the
flavor mixing matrix. Some phenomenological implications of such
democratic quark mass matrices, including their variations in the
hierarchy basis and their evolution from the electroweak scale to a
superhigh-energy scale, are also discussed.
\end{abstract}

\begin{flushleft}
\hspace{0.8cm} PACS number(s): 12.15.Ff, 12.39.-x, 11.10.Hi
\end{flushleft}

\def\thefootnote{\arabic{footnote}}
\setcounter{footnote}{0}

\newpage

\section{Introduction}

In the standard electroweak model the origin of quark masses is
attributed to the Yukawa interactions and the Higgs mechanism. But
the model gives no quantitative prediction for the structures of the
Yukawa coupling matrices $Y^{}_{+2/3}$ and $Y^{}_{-1/3}$ in the
$Q=+2/3$ and $Q=-1/3$ quark sectors, respectively. That is why there
is no explanation of the observed strong hierarchies of quark
masses, namely $m^{}_u/m^{}_c \sim m^{}_c/m^{}_t \sim \lambda^4$ and
$m^{}_d/m^{}_s \sim m^{}_s/m^{}_b \sim \lambda^2$ with $\lambda
\simeq 0.2$ \cite{XZZ}, within the standard model. In other words,
why the three eigenvalues of the Yukawa coupling matrix
$Y^{}_{+2/3}$ or $Y^{}_{-1/3}$ (i.e., $f^{}_\alpha = m^{}_\alpha/v$
with $v \simeq 174$ GeV being the vacuum expectation value and
$\alpha$ running over $u$, $c$ and $t$ for $Y^{}_{+2/3}$ or $d$, $s$
and $b$ for $Y^{}_{-1/3}$) are so different in magnitude? This
remains a highly puzzling question.

As first pointed out by Harari, Haut and Weyers in 1978 \cite{Harari},
it should be very natural to conjecture that the quark fields of the
same electric charge initially have the identical Yukawa
interactions with the Higgs field, namely,
\begin{eqnarray}
Y^{(0)}_Q = \frac{C^{(0)}_Q}{3} \left( \begin{matrix} 1 & 1 & 1 \cr
1 & 1 & 1 \cr 1 & 1 & 1 \cr
\end{matrix} \right) \; ,
%     (1)
\end{eqnarray}
where $C^{(0)}_Q$ is a dimensionless coefficient, and $Q = +2/3$ for
the up-quark sector or $Q= -1/3$ for the down-quark sector. Such a
form of $Y^{(0)}_Q$ means that the corresponding quark mass matrix
$M^{(0)}_Q$ must have the same ``flavor democracy",
\begin{eqnarray}
M^{(0)}_Q = \frac{m^{}_3}{3} \left( \begin{matrix} 1 & 1 & 1 \cr 1 &
1 & 1 \cr 1 & 1 & 1 \cr
\end{matrix} \right) \; ,
%     (2)
\end{eqnarray}
where $m^{}_3 \equiv v C^{(0)}_Q$, equal to the top-quark mass
$m^{}_t$ for $Q= +2/3$ or the bottom-quark mass $m^{}_b$ for $Q=
-1/3$. The corresponding quark mass term can be written as
\begin{eqnarray}
\frac{m^{}_3}{3} \sum_\alpha \sum_\beta \overline{\alpha^{}_{\rm L}}
\ \beta^{}_{\rm R} + {\rm h.c.} \; ,
%     (3)
\end{eqnarray}
and it is completely invariant under the permutation of all the
three left-handed quark fields and all the three right-handed quark
fields, where $\alpha, \beta = u, c, t$ for $Q=+2/3$ or $\alpha,
\beta = d, s, b$ for $Q=-1/3$. That is to say, the flavor democracy
of $Y^{(0)}_Q$ or $M^{(0)}_Q$ implies that the quark mass term in
Eq. (3) possesses the exact $S(3)^{}_{\rm L} \times S(3)^{}_{\rm R}$
symmetry. This symmetry must be broken, since two of the three
eigenvalues of $M^{}_Q$ are vanishing. The breaking of this flavor
democracy leads to the flavor mixing effects between the two quark
sectors \cite{D1,Yang,D2}.

How to break the democracy of quark flavors and to what extent to
break it are two highly nontrivial questions for model building in
this regard \cite{FX2000}. In the present work we are going to
address ourselves to these two questions by assuming a structural
parallelism between the mass matrices of $Q=+2/3$ and $Q=-1/3$
quarks. Such a phenomenological assumption makes sense if the
generation of quark masses in the two sectors is governed by the
same dynamics, and combining it with a nontrivial parametrization of
the Cabibbo-Kobayashi-Maskawa (CKM) quark mixing matrix proposed by
Fritzsch and Xing \cite{FX} allows one to figure out the texture and
strength of flavor democracy breaking in each quark sector in terms
of the observed values of quark masses and flavor mixing parameters.
Some interesting implications of such flavor-democratized quark mass
matrices, including their variations in the hierarchy basis and
their evolution with the energy scales, are also discussed.

\section{Flavor democracy breaking}

Let us begin with diagonalizing the flavor-democratized
quark mass matrix $M^{(0)}_Q$ as follows:
\begin{eqnarray}
V^\dagger_0 M^{(0)}_Q V^{}_0 = m^{}_3 \left(\begin{matrix} 0 & 0 & 0
\cr 0 & 0 & 0 \cr 0 & 0 & 1 \cr\end{matrix}\right) \; ,
%     (4)
\end{eqnarray}
where
\begin{eqnarray}
V^{}_0 = \left(\begin{matrix}
\frac{1}{\sqrt{2}} & \frac{1}{\sqrt{6}} & \frac{1}{\sqrt{3}}
\cr -\frac{1}{\sqrt{2}} & \frac{1}{\sqrt{6}} & \frac{1}{\sqrt{3}} \cr
0 & -\frac{2}{\sqrt{6}} & \frac{1}{\sqrt{3}} \cr\end{matrix}\right) \; .
%     (5)
\end{eqnarray}
We therefore arrive at $m^{}_1 = m^{}_2 =0$, which are qualitatively
consistent with the experimental fact $m^{}_u, m^{}_c \ll m^{}_t$ or
$m^{}_d, m^{}_s \ll m^{}_b$. However, there is no flavor mixing in
this special case, because the resulting CKM matrix $V = V^\dagger_0
V^{}_0 = {\bf 1}$ is an identity matrix.

The realistic CKM quark mixing matrix
\begin{eqnarray}
V = V^\dagger_{+2/3} V^{}_{-1/3} = (V^{}_0 V^{}_{+2/3})^\dagger
(V^{}_0 V^{}_{-1/3})
%     (6)
\end{eqnarray}
measures a mismatch between the diagonalization of the $Q=+2/3$
quark mass matrix $M^{}_{+2/3}$ and that of the $Q=-1/3$ quark mass
matrix $M^{}_{-1/3}$, and thus it provides a natural description of
the observed phenomena of quark flavor mixing. Notice that
$M^{}_{+2/3}$ and $M^{}_{-1/3}$ can always be arranged to be
Hermitian, thanks to a proper choice of the flavor basis in the
standard model or its extensions which have no flavor-changing
right-handed currents \cite{Frampton}. So let us simply focus on
Hermitian quark mass matrices in the following and take into account
the corresponding flavor democracy in such a basis, namely,
\begin{eqnarray}
(V^{}_0 V^{}_{Q})^\dagger M^{}_{Q} (V^{}_0 V^{}_{Q}) =
\widehat{M}^{}_{Q} \equiv \left(\begin{matrix} m^{}_1 & 0 & 0 \cr 0
& m^{}_2 & 0 \cr 0 & 0 & m^{}_3 \cr\end{matrix} \right) \; ,
%     (7)
\end{eqnarray}
where $m^{}_1 = \pm m^{}_u$, $m^{}_2 = \pm m^{}_c$ and $m^{}_3 =
m^{}_t$ for $Q=+2/3$, or $m^{}_1 = \pm m^{}_d$, $m^{}_2 = \pm
m^{}_s$ and $m^{}_3 = m^{}_b$ for $Q=-1/3$. Here the sign ambiguity
of $m^{}_1$ or $m^{}_2$ is attributed to the fact that the
eigenvalues of the Hermitian matrix $M^{}_Q$ can be either positive
or negative under the above unitary transformation. To reconstruct
the pattern of $M^{}_Q$ in terms of $V^{}_0$, $V^{}_Q$ and
$\widehat{M}^{}_Q$, however, one must specify the form of $V^{}_Q$
with the help of the parameters of $V$.

We find that the most suitable parametrization of the CKM matrix $V$
for our purpose is the one advocated by two of us in Ref. \cite{FX}:
\begin{eqnarray}
V=\left( \begin{matrix} \sin\theta^{}_{\rm u}\sin\theta^{}_{\rm d}
\cos\theta + \cos\theta^{}_{\rm u}
\cos\theta^{}_{\rm d} e^{-{\rm i}\phi} & \sin\theta^{}_{\rm u}
\cos\theta^{}_{\rm d} \cos\theta - \cos\theta^{}_{\rm u}
\sin\theta^{}_{\rm d} e^{-{\rm i}\phi} & \sin\theta^{}_{\rm u}
\sin\theta \cr
\cos\theta^{}_{\rm u} \sin\theta^{}_{\rm d} \cos\theta -
\sin\theta^{}_{\rm u} \cos\theta^{}_{\rm d} e^{-{\rm i}\phi}
& \cos\theta^{}_{\rm u} \cos\theta^{}_{\rm d} \cos\theta +
\sin\theta^{}_{\rm u} \sin\theta^{}_{\rm d} e^{-{\rm i}\phi}
& \cos\theta^{}_{\rm u} \sin\theta \cr -\sin\theta^{}_{\rm d}
\sin\theta & -\cos\theta^{}_{\rm d} \sin\theta & \cos\theta
\end{matrix} \right) \;
%     (8)
\end{eqnarray}
with the subscripts ``u" and ``d" denoting ``up" ($Q=+2/3$) and
``down" ($Q=-1/3$), respectively. The reason is simply that this
form of $V$ can be decomposed into $V^{}_{+2/3}$ and $V^{}_{-1/3}$
in an exactly {\it parallel} way as follows:
\begin{eqnarray}
V^{}_{+2/3} = \left(\begin{matrix} 1 & 0 & 0 \cr 0 &
\cos (+\frac{2}{3}\theta) & -\sin (+\frac{2}{3}\theta) \cr
0 & \sin (+\frac{2}{3}\theta) & \cos (+\frac{2}{3}\theta)
\end{matrix}\right) \left(\begin{matrix}
\exp (+{\rm i} \frac{2}{3}\phi) & 0 & 0 \cr 0 & 1 & 0
\cr 0 & 0 & 1 \end{matrix}\right)
\left(\begin{matrix} \cos\theta^{}_{\rm u} & -\sin\theta^{}_{\rm u}
& 0 \cr \sin\theta^{}_{\rm u} & \cos\theta^{}_{\rm u} & 0
\cr 0 & 0 & 1\end{matrix}\right) \; ,
\nonumber \\
V^{}_{-1/3} = \left(\begin{matrix} 1 & 0 & 0 \cr 0 & \cos
(-\frac{1}{3}\theta) & -\sin (-\frac{1}{3}\theta) \cr 0 & \sin
(-\frac{1}{3}\theta) & \cos (-\frac{1}{3}\theta) \end{matrix}\right)
\left(\begin{matrix} \exp(-{\rm i} \frac{1}{3}\phi) & 0 & 0 \cr 0 &
1 & 0 \cr 0 & 0 & 1 \end{matrix}\right) \left(\begin{matrix}
\cos\theta^{}_{\rm d} & -\sin\theta^{}_{\rm d} & 0 \cr
\sin\theta^{}_{\rm d} & \cos\theta^{}_{\rm d} & 0 \cr 0 & 0 & 1
\end{matrix}\right) \; .
%     (9)
\end{eqnarray}
Since all the four parameters in this parametrization of $V$ can be
determined to a good degree of accuracy by using current
experimental data, one may therefore fix the patterns of
$V^{}_{+2/3}$ and $V^{}_{-1/3}$. Of course, the decomposition made
in Eq. (9) depends also on a purely phenomenological assumption: the
up- and down-type components of the flavor mixing angle $\theta$ are
demanded to be proportional to the corresponding charges of these
two quark sectors, so are the components of the CP-violating phase
$\phi$. Such an assumption is another reflection of the {\it up-down
parallelism}, which has been taken as the main guiding principle of
our treatment, although it is very hard to argue any potential
connection between the quark mass textures and the quark charges at
this stage
%%%%%%%%%%%%%%%%%%%%%%%%%%%%%%
\footnote{However, it has been argued that the origin of some
differences between the up- and down-quark sectors might simply
represent a difference between their charges in a dynamical model
which can explain the observed family structure, rather than a
fundamental difference between the two sectors \cite{Hung}.}.
%%%%%%%%%%%%%%%%%%%%%%%%%%%%%%
One is certainly allowed to try some other possibilities of
decomposing $V$ into $V^{}_{+2/3}$ and $V^{}_{-1/3}$ \cite{D2}, but
the key point should be the same as ours --- to minimize the number
of free parameters in reason, at least at the phenomenological
level.

Given Eqs. (7) and (9), we are now in a position to reconstruct the
quark mass matrices $M^{}_{+2/3}$ and $M^{}_{-1/3}$ based on the
flavor democracy. The texture of $M^{}_Q$ can be expressed as
\begin{eqnarray}
M^{}_{Q} = A^{2}_Q M^{(0)}_{Q} + M^{(1)}_{Q} + M^{(2)}_{Q} \; ,
%     (10)
\end{eqnarray}
where $A^{}_{Q} = -\sin{(Q\theta)}/\sqrt{2}+\cos{(Q\theta)}$,
$M^{(0)}_Q$ has been defined in Eq. (2), and
\begin{eqnarray}
\begin{aligned}
M^{(1)}_{Q}&=C^{(11)}_{Q}\left( \begin{matrix} 1 & 1 & -r^{}_Q \cr 1 & 1
& -r^{}_Q  \cr -r^{}_Q & -r^{}_Q  & r^{2}_Q
\end{matrix} \right)+C^{(12)}_{Q}\left( \begin{matrix} 0 & 0 &
r^{}_Q \cr 0 & 0 & r^{}_Q  \cr r^{}_Q  & r^{}_Q  & 2+r^{}_Q
\end{matrix} \right)  \; ,
\\
M^{(2)}_{Q}&=C^{(21)}_{Q}\left[\cos{(Q\phi)}\left(\begin{matrix} 1 &
0 & -1\cr 0 & -1 & 1 \cr -1 & 1 & 0 \end{matrix}\right)+{\rm i}
\sin{(Q\phi)} \left(\begin{matrix} 0 & 1 & -1\cr -1 & 0 & 1 \cr 1 &
-1 & 0 \end{matrix} \right) \right]
\\
&+C^{(22)}_{Q}\left(\begin{matrix} 1 & -1 & 0 \cr -1 & 1 & 0 \cr
0 & 0 & 0 \end{matrix} \right)
\\
&+C^{(23)}_{Q}\left[\cos{(Q\phi)} \left(\begin{matrix} 2 & 0 & 1\cr
0 & -2 & -1 \cr 1 & -1 & 0 \end{matrix}\right)-{\rm
i}\sin{(Q\phi)}\left(\begin{matrix} 0 & -2 & -1\cr 2 & 0 & 1 \cr 1 &
-1 & 0 \end{matrix}\right) \right] \;,
\end{aligned}
%   (11)
\end{eqnarray}
in which $r^{}_Q = 2A^{}_Q/B^{}_Q$ with $B^{}_{Q}=\sqrt{2}
\sin{(Q\theta)}+\cos{(Q\theta)}$, and
\begin{equation}
\begin{aligned}
& C^{(11)}_{Q} = \frac{1}{6}\left(m^{}_{1}\sin^{2}{\theta^{}_{\rm q}}
+ m^{}_{2}\cos^{2} {\theta^{}_{\rm q}}\right) B^{2}_{Q} \; ,
\\
& C^{(12)}_{Q} = \frac{1}{2\sqrt{2}}m^{}_{3}\sin{(Q\theta)}B^{}_{Q} \; ,
\\
& C^{(21)}_{Q} = \frac{1}{2\sqrt{3}}(m^{}_{1}-m^{}_{2})\cos{(Q\theta)}
\sin{2\theta^{}_{\rm q}} \; ,
\\
& C^{(22)}_{Q} = \frac{1}{2}\left(m^{}_{1}\cos^{2}{\theta^{}_{\rm q}}
+ m^{}_{2}\sin^{2}{\theta^{}_{\rm q}}\right) \; ,
\\
& C^{(23)}_{Q} = \frac{1}{2\sqrt{6}}(m^{}_{1}-m^{}_{2})\sin{(Q\theta)}
\sin{2\theta^{}_{\rm q}} \;
\end{aligned}
%     (12)
\end{equation}
with $\rm q=u$ for $Q=+2/3$ and $\rm q=d$ for $Q=-1/3$. It is
obvious that the matrices $M^{(0)}_{Q}$, $M^{(1)}_{Q}$ and
$M^{(2)}_{Q}$ perform the $S(3)^{}_{\rm L}\times S(3)^{}_{\rm R}$,
$S(2)^{}_{\rm L}\times S(2)^{}_{\rm R}$ and $S(1)^{}_{\rm L}\times
S(1)^{}_{\rm R}$ flavor symmetries, respectively. Among the five
coefficients of $M^{(1)}_{Q}$ and $M^{(2)}_{Q}$ in Eq. (12),
$C^{(12)}_{Q}$ is proportional to $m^{}_3 \sin(Q\theta)$ and the
others are all dominated by the terms proportional to $m^{}_2$.
Hence their ratios to the coefficient of $M^{(0)}_Q$ (i.e.,
$m^{}_3/3$) are suppressed at the levels of $\sin(Q\theta)$ and
$m^{}_2/m^{}_3$, respectively. Because $\theta \sim \lambda^2$
\cite{FX} and $|m^{}_2/m^{}_3| \sim \lambda^4$ (for $Q=+2/3$) or
$\lambda^2$ (for $Q=-1/3$) \cite{XZZ}, the relevant suppression is
at least at the percent level. In other words, the strength of
flavor democracy breaking must be at or below the percent level.

To see this point more clearly, let us take account of the strong
quark mass hierarchy and the smallness of three flavor mixing angles
to make a reasonable analytical approximation for the expression of
$M^{}_Q$ in Eq. (10). Then we arrive at
\begin{eqnarray}
\begin{aligned}
M^{}_{Q} & \simeq \frac{1}{3}m^{}_{3}\left\{\left( \begin{matrix} 1
& 1 & 1 \cr 1 & 1 & 1 \cr 1 & 1 & 1
\end{matrix}\right)\right.
+\left[\frac{1}{2}\frac{m^{}_{2}}{m^{}_{3}}\left( \begin{matrix} 1 &
1 & -r \cr 1 & 1 & -r \cr -r & -r & r^2
\end{matrix} \right) +\frac{3\sqrt{2}}{4}Q\theta\left(
\begin{matrix} 0 & 0 & r \cr 0 & 0 & r \cr r &
r & 2+r \end{matrix}
\right)\right]
\\
&-\sqrt{3} \ \theta^{}_{\rm
q}\frac{m^{}_{2}}{m^{}_{3}}\left[\cos{(Q\phi)} \left(\begin{matrix}1
& 0 & -1\cr 0 & -1 & 1 \cr -1 & 1 & 0
\end{matrix}\right)+{\rm i}\sin{(Q\phi)} \left(\begin{matrix} 0 & 1
& -1\cr -1 & 0 & 1 \cr 1 & -1 & 0 \end{matrix} \right) \right]
\\
&-\frac{\sqrt{6}}{2}Q\theta\theta^{}_{\rm
q}\frac{m^{}_{2}}{m^{}_{3}} \left[\cos{(Q\phi)}\left(\begin{matrix}
2 & 0 & 1\cr 0& -2 & -1 \cr 1 & -1 & 0 \end{matrix}\right)-{\rm
i}\sin{(Q\phi)}\left(\begin{matrix} 0 & -2 & -1\cr 2 & 0 & 1 \cr 1 &
-1 & 0 \end{matrix}\right)\right]
\\
&\left.+\frac{3}{2}\left(\frac{m^{}_{1}}{m^{}_{3}}+\theta^{2}_{\rm
q}\frac{m^{}_{2}} {m^{}_{3}}\right)\left(\begin{matrix} 1 & -1 &
0\cr -1 & 1 & 0 \cr 0 & 0 & 0 \end{matrix} \right)\right\} \;,
\end{aligned}
%   (13)
\end{eqnarray}
in which the subscript of $r^{}_Q$ has been omitted. In fact,
$r^{}_Q \simeq 2-3\sqrt{2} \ Q\theta$ is not very sensitive to the
value of $Q$ due to the smallness of $\theta$. The result in Eq.
(13) shows a hierarchical chain of flavor democracy breaking in the
quark sector. First, the $S(3)^{}_{\rm L}\times S(3)^{}_{\rm R}$
symmetry is broken down to the $S(2)^{}_{\rm L}\times S(2)^{}_{\rm
R}$ symmetry, and the strength of this effect is characterized by
the small quantities $m^{}_2/m^{}_3$ and $\theta$. Second, the
$S(2)^{}_{\rm L}\times S(2)^{}_{\rm R}$ symmetry is further broken
down to $S(1)^{}_{\rm L}\times S(1)^{}_{\rm R}$, and the
corresponding effect is further suppressed because it is
characterized by the much smaller quantities $\theta^{}_{\rm q}
m^{}_2/m^{}_3$, $\theta \theta^{}_{\rm q} m^{}_2/m^{}_3$,
$\theta^{2}_{\rm q} m^{}_2/m^{}_3$ and $m^{}_1/m^{}_3$. In
particular, the CP-violating phase $\phi$ comes in at the second
symmetry-breaking stage and hence the effect of CP violation is
strongly suppressed.

We proceed to evaluate the strength of flavor democracy breaking in
a numerical way. To do so, we make use of the central values of six
quark masses renormalized to the electroweak scale characterized by
the $Z$-boson mass \cite{XZZ}:
\begin{eqnarray}
\begin{aligned}
& m^{}_{u} \simeq 1.38 ~{\rm MeV} \; , ~~~ m^{}_{c} \simeq 638 ~
{\rm MeV} \; , ~~~ m^{}_t \simeq 172.1 ~{\rm GeV} \; ;
\\
& m^{}_{d} \simeq 2.82 ~{\rm MeV} \; , ~~~ m^{}_s \simeq 57 ~{\rm
MeV} \; , ~~~ m^{}_{b} \simeq 2.86 ~{\rm GeV} \; .
\end{aligned}
%   (14)
\end{eqnarray}
The values of the flavor mixing parameters $\theta^{}_{\rm u}$,
$\theta^{}_{\rm d}$, $\theta$ and $\phi$ can be obtained by
establishing their relations with the well-known Wolfenstein
parameters \cite{L}, whose values have been determined to an
impressively good degree of accuracy \cite{PDG,CKM}:
\begin{eqnarray}
\begin{aligned}
&\theta^{}_{\rm u} \simeq
\arctan{\left(\lambda\sqrt{\overline{\rho}^{2}+\overline{\eta}^{2}}
\right)} \simeq 0.086 \; ,
\\
&\theta^{}_{\rm d} \simeq
\arctan{\left(2\lambda\sqrt{\frac{(1-\overline{\rho})^{2}+
\overline{\eta}^{2}}{\left[\lambda^{2}(1-2\overline{\rho})-2\right]^{2}
+4\lambda^{4} \overline{\eta}^2}}\right)} \simeq 0.206 \; ,
\\
&\theta \simeq \arcsin{\left(A\lambda^{2}\sqrt{1+\lambda^{2}
\left(\overline{\rho}^2+ \overline{\eta}^2\right)}\right)} \simeq
0.042 \; ,
\\
&\phi \simeq \arccos{\left(\frac{\sin^2{\theta^{}_{\rm
u}}\cos^2{\theta^{}_{\rm d}}\cos^2{\theta} + \cos^2{\theta{}_{\rm
u}}\sin^2{\theta^{}_{\rm d}} - \lambda^{2}}{2\sin{\theta^{}_{\rm u}}
\cos{\theta^{}_{\rm u}}\sin{\theta^{}_{\rm d}}\cos{\theta^{}_{\rm
d}}\cos{\theta}}\right)} \simeq 1.636 \; ,
%   (15)
\end{aligned}
\end{eqnarray}
where the best-fit values $A \simeq 0.825$, $\lambda \simeq 0.2251$,
$\overline{\rho} \simeq 0.160$ and $\overline{\eta} \simeq 0.350$
\cite{CKM} have been input. Namely, we have
\begin{eqnarray}
\theta^{}_{\rm u} \simeq 4.951^{\circ} \; , ~~~ \theta^{}_{\rm d}
\simeq 11.772^{\circ} \; , ~~~ \theta \simeq 2.405^{\circ} \; , ~~~
\phi \simeq 93.730^{\circ} \; ,
%     (16)
\end{eqnarray}
implying $\theta^{}_{\rm u} \sim 2\lambda^2$, $\theta^{}_{\rm d}
\sim \lambda$ and $\theta \sim \lambda^2$ in terms of the expansion
parameter $\lambda \simeq 0.2$. The fact that $\phi$ is very close
to $\pi/2$ proves to be quite suggestive in quark flavor
phenomenology, as already discussed in Ref. \cite{LX}.

With the help of the central values of six quark masses and four
flavor mixing parameters given in Eqs. (14) and (16), one may start
from Eq. (10) to numerically calculate the elements of $M^{}_{+2/3}$
and $M^{}_{-1/3}$ in two typical possibilities:

(a) $(m^{}_{1}, m^{}_2) = (-m^{}_u, +m^{}_c)$ for $Q=+2/3$ and
$(-m^{}_d, +m^{}_s)$ for $Q=-1/3$, leading to
\begin{eqnarray}
\begin{aligned}
M^{}_{+2/3} & \simeq 55.07 ~{\rm GeV} \times \left\{\left(
\begin{matrix} 1 & 1 & 1 \cr 1 & 1 & 1\cr 1 & 1 & 1
\end{matrix}\right) + \left[3.2\times10^{-2}\left( \begin{matrix} 0 &
0 & 1.89 \cr 0 & 0 & 1.89 \cr 1.89 & 1.89 & 3.89 \end{matrix}\right)
\right.\right.
\\
& \left. -2.07 \times 10^{-3}\left( \begin{matrix} -1 & -1 & 1.89
\cr -1 & -1 & 1.89 \cr 1.89 & 1.89 & -3.56
\end{matrix}\right)\right] -\left[ 2.65 \times
10^{-4}\left(\begin{matrix} 1 & 0 & -1 \cr 0 & -1 & 1 \cr -1 & 1 & 0
\end{matrix}\right) \right.
\\
& \left. -3.03 \times 10^{-5}\left(\begin{matrix} 1 & -1 & 0 \cr -1
& 1 & 0 \cr 0 & 0 & 0 \end{matrix}\right)+5.24 \times
10^{-6}\left(\begin{matrix} 2 & 0 & 1 \cr 0 & -2 & -1 \cr 1 & -1 & 0
\end{matrix}\right) \right]
\\
& \left. - {\rm i} \left[5.09 \times 10^{-4} \left(\begin{matrix} 0
& 1 & -1 \cr -1 & 0 & 1\cr 1 & -1 & 0 \end{matrix}\right) -1.01
\times 10^{-5} \left(\begin{matrix} 0 & -2 & -1 \cr 2 & 0 & 1 \cr1 &
-1 & 0 \end{matrix}\right) \right]\right\} \; ,
%   (17)
\end{aligned}
\end{eqnarray}
and
\begin{eqnarray}
\begin{aligned}
M_{-1/3} & \simeq 0.97 ~{\rm GeV} \times \left\{\left(\begin{matrix}
1 & 1 & 1 \cr 1 & 1 & 1 \cr 1 & 1 & 1
\end{matrix}\right)-\left[1.43 \times 10^{-2}\left(\begin{matrix} 0 &
0 & 2.06 \cr 0 & 0 & 2.06 \cr 2.06 & 2.06 & 4.06 \end{matrix}\right)
\right.\right.
\\
&\left.+8.98 \times 10^{-3}\left(\begin{matrix} -1 & -1 & 2.06 \cr
-1 & -1 & 2.06 \cr 2.06 & 2.06 & -4.25
\end{matrix}\right)\right]-\left[ 6.08 \times
10^{-3}\left(\begin{matrix} 1 & 0 & -1 \cr 0 & -1 & 1 \cr -1 & 1 & 0
\end{matrix}\right) \right.
\\
&\left. +1.63 \times 10^{-4}\left(\begin{matrix} 1 & -1 & 0 \cr -1 &
1 & 0 \cr 0 & 0 & 0 \end{matrix}\right)-6.02 \times 10^{-5}
\left(\begin{matrix} 2 & 0 & 1 \cr 0 & -2 & -1 \cr 1 & -1 & 0
\end{matrix}\right)\right]
\\
&+\left. {\rm i} \left[ 3.69 \times 10^{-3} \left(\begin{matrix} 0 &
1 & -1 \cr -1 & 0 & 1 \cr 1 & -1 & 0 \end{matrix}\right)+3.65 \times
10^{-5} \left(\begin{matrix} 0 & -2 & -1 \cr 2 & 0 & 1 \cr 1 & -1 &
0 \end{matrix} \right) \right] \right\} \; ;
%   (18)
\end{aligned}
\end{eqnarray}

(b) $(m^{}_{1}, m^{}_2) = (+m^{}_u, +m^{}_c)$ for $Q=+2/3$ and
$(+m^{}_d, +m^{}_s)$ for $Q=-1/3$, leading to
\begin{eqnarray}
\begin{aligned}
M^{}_{+2/3}& \simeq 55.07 ~{\rm GeV} \times\left\{\left(
\begin{matrix} 1 & 1 & 1 \cr 1 & 1 & 1 \cr 1 & 1 & 1
\end{matrix}\right) +\left[ 3.2 \times 10^{-2}\left(\begin{matrix} 0
& 0 & 1.89 \cr 0 & 0 & 1.89 \cr 1.89 & 1.89 & 3.89
\end{matrix}\right) \right.\right.
\\
& \left. -2.07 \times 10^{-3}\left(\begin{matrix} -1 & -1 & 1.89 \cr
-1 & -1 & 1.89 \cr 1.89 & 1.89 & -3.56 \end{matrix} \right)
\right]-\left[ 2.64 \times 10^{-4}\left(\begin{matrix} 1 & 0 & -1
\cr 0 & -1 & 1 \cr -1 & 1 & 0 \end{matrix}\right) \right.
\\
& \left. -5.52 \times 10^{-5}\left(\begin{matrix} 1 & -1 & 0 \cr -1
& 1 & 0 \cr 0 & 0 & 0 \end{matrix}\right) + 5.22 \times
10^{-6}\left(\begin{matrix} 2 & 0 & 1 \cr 0 & -2 & -1 \cr 1 & -1 & 0
\end{matrix}\right) \right]
\\
&-\left. {\rm i} \left[ 5.06 \times 10^{-4} \left(\begin{matrix} 0 &
1 & -1 \cr -1 & 0 & 1 \cr 1 & -1 & 0 \end{matrix}\right) - 1.00
\times 10^{-5} \left( \begin{matrix} 0 & -2 & -1 \cr 2 & 0 & 1 \cr1
& -1 & 0 \end{matrix} \right) \right] \right\}\;,
%   (19)
\end{aligned}
\end{eqnarray}
and
\begin{eqnarray}
\begin{aligned}
M_{-1/3}& \simeq 0.97 ~{\rm GeV} \times \left\{ \left(\begin{matrix}
1 & 1 & 1 \cr 1 & 1 & 1 \cr 1 & 1 & 1 \end{matrix}\right) - \left[
1.43 \times 10^{-2}\left(\begin{matrix} 0 & 0 & 2.06 \cr 0 & 0 &
2.06 \cr 2.06 & 2.06 & 4.06 \end{matrix}\right)\right. \right.
\\
& \left. +9.01 \times 10^{-3} \left(\begin{matrix} -1 & -1 & 2.06
\cr -1 & -1 & 2.06 \cr 2.06 & 2.06 & -4.25
\end{matrix}\right)\right] - \left[5.51 \times 10^{-3}
\left(\begin{matrix} 1 & 0 & -1 \cr 0 & -1 & 1 \cr -1 & 1 & 0
\end{matrix}\right) \right.
\\
&\left.-2.62 \times 10^{-3} \left(\begin{matrix} 1 & -1 & 0 \cr -1 &
1 & 0 \cr 0 & 0 & 0 \end{matrix}\right)-5.45 \times
10^{-5}\left(\begin{matrix} 2 & 0 & 1 \cr 0 & -2 & -1 \cr 1 & -1 & 0
\end{matrix}\right)\right]
\\
&+\left. {\rm i} \left[3.34 \times 10^{-3} \left(\begin{matrix} 0 &
1 & -1 \cr -1 & 0 & 1 \cr 1 & -1 & 0 \end{matrix} \right) + 3.31
\times 10^{-5} \left(\begin{matrix} 0 & -2 & -1 \cr 2 & 0 & 1 \cr 1
& -1 & 0 \end{matrix} \right) \right] \right\}\;.
%   (20)
\end{aligned}
\end{eqnarray}
Some comments on the implications of these results are in order.
\begin{itemize}
\item     The other two possibilities, corresponding to $(m^{}_1,
m^{}_2) = (+m^{}_u, -m^{}_c)$ and $(-m^{}_u, -m^{}_c)$ in the
$Q=+2/3$ quark sector or $(m^{}_1, m^{}_2) = (+m^{}_d, -m^{}_s)$ and
$(-m^{}_d, -m^{}_s)$ in the $Q=-1/3$ quark sector, are numerically
found to be very similar to cases (a) and (b) shown above. Hence
they will not be separately discussed.

\item     The $S(2)^{}_{\rm L} \times S(2)^{}_{\rm R}$ terms of
$M^{}_Q$ are not sensitive to the sign ambiguities of $m^{}_1$ and
$m^{}_2$, but the latter can affect those $S(1)^{}_{\rm L} \times
S(1)^{}_{\rm R}$ terms of $M^{}_Q$ to some extent. In other words, a
specific model-building exercise should take into account the fine
structure of $M^{}_Q$ which is associated with both the lightest
quark mass and the CP-violating phase in each quark sector.

\item     It is always possible to combine the two
$S(2)^{}_{\rm L} \times S(2)^{}_{\rm R}$ terms of $M^{}_Q$, and such
a combination does not violate the $S(2)^{}_{\rm L} \times
S(2)^{}_{\rm R}$ symmetry. Since the coefficients of five
$S(1)^{}_{\rm L} \times S(1)^{}_{\rm R}$ terms are very different in
magnitude, it is reasonable to neglect the most strongly suppressed
ones when building a phenomenologically viable quark mass model. In
particular, Eqs. (17)---(20) suggest that $C^{(22)}_Q \simeq 0$ and
$C^{(23)}_Q \simeq 0$ should be two good approximations, which can
also be observed from their analytical expressions in Eq. (12) or
(13) by considering $|m^{}_1| \ll |m^{}_2| \ll m^{}_3$ and the smallness
of $\theta$ and $\theta^{}_{\rm q}$. In this situation the
analytical approximation of $M^{}_Q$ in Eq. (13) is further
simplified to
\begin{eqnarray}
\begin{aligned}
M^{}_{Q} & \simeq \frac{1}{3}m^{}_{3}\left\{\left( \begin{matrix} 1
& 1 & 1 \cr 1 & 1 & 1 \cr 1 & 1 & 1
\end{matrix}\right)\right.
+\left[\frac{1}{2}\frac{m^{}_{2}}{m^{}_{3}}\left( \begin{matrix} 1 &
1 & -2 \cr 1 & 1 & -2 \cr -2 & -2 & 4
\end{matrix} \right) +\frac{3\sqrt{2}}{4}Q\theta\left(
\begin{matrix} 0 & 0 & 2 \cr 0 & 0 & 2 \cr 2 &
2 & 4 \end{matrix} \right)\right]
\\
& \left. -\sqrt{3} \ \theta^{}_{\rm
q}\frac{m^{}_{2}}{m^{}_{3}}\left[\cos{(Q\phi)} \left(\begin{matrix}1
& 0 & -1\cr 0 & -1 & 1 \cr -1 & 1 & 0
\end{matrix}\right)+{\rm i}\sin{(Q\phi)} \left(\begin{matrix} 0 & 1
& -1\cr -1 & 0 & 1 \cr 1 & -1 & 0 \end{matrix} \right)
\right]\right\} \;,
\end{aligned}
%   (21)
\end{eqnarray}
where $r\simeq 2$ has been taken into account.
\end{itemize}
In short, the strength of $S(3)^{}_{\rm L} \times S(3)^{}_{\rm R}
\to S(2)^{}_{\rm L} \times S(2)^{}_{\rm R}$ breaking is at the
percent level for both up- and down-quark sectors, while the effects
of $S(2)^{}_{\rm L} \times S(2)^{}_{\rm R} \to S(1)^{}_{\rm L}
\times S(1)^{}_{\rm R}$ breaking are at the percent and ten percent
levels for the up- and down-quark sectors, respectively.

\section{On the hierarchy basis}

It is sometimes convenient to ascribe the hierarchy of the quark
mass spectrum directly to the hierarchy of the corresponding quark
mass matrix. In the latter basis, which is usually referred to as
the hierarchy basis, the quark mass matrix $M^\prime_Q$ is related
to its democratic counterpart $M^{}_Q$ via the following
transformation:
\begin{eqnarray}
M^\prime_Q = V^{\dag}_{0}M^{}_Q V^{}_{0} \; ,
%   (22)
\end{eqnarray}
where $V^{}_0$ and $M^{}_Q$ have been given in Eqs. (5) and (10),
respectively. To be explicit, we obtain
\begin{eqnarray}
M^\prime_Q = \left(\begin{matrix} 2C^{(22)}_Q & \sqrt{3} \
C^{(21)}_Q e^{{\rm i}Q\phi} & \sqrt{6} \ C^{(23)}_Q e^{{\rm i}Q\phi} \cr
\sqrt{3} \ C^{(21)}_Q e^{-{\rm i}Q\phi} &
X^{}_Q & Y^{}_Q \cr
\sqrt{6} \ C^{(23)}_Q e^{-{\rm i}Q\phi} &
Y^{}_Q & Z^{}_Q \end{matrix}\right)  \; ,
%   (23)
\end{eqnarray}
where
\begin{eqnarray}
&& X^{}_Q = \frac{2}{3}\left[\left(r^{}_Q+1\right)^2 C^{(11)}_Q -
\left(r^{}_Q-2\right) C^{(12)}_Q \right] \; ,
\nonumber \\
&& Y^{}_Q = -\frac{\sqrt{2}}{3}\left(r^{}_Q+1\right)\left[\left(r^{}_Q-2
\right) C^{(11)}_Q +2C^{(12)}_Q \right] \; ,
\nonumber \\
&& Z^{}_Q = \frac{1}{3}\left[\left(r^{}_Q-2\right)^{2}C^{(11)}_Q +
\left(5 \ r^{}_Q + 2\right)C^{(12)}_Q \right] + A^{2}_Q m^{}_{3} \; .
%     (24)
\end{eqnarray}
The exact expression of $M^\prime_Q$ in Eq. (23) can be simplified,
if the analytical approximation made in Eq. (13) for $M^{}_Q$ is
taken into account. In this case,
\begin{eqnarray}
M^\prime_Q \simeq \left(\begin{matrix} m^{}_{1}+\theta^{2}_{\rm
q}m^{}_{2} & -\theta^{}_{\rm q} m^{}_{2} e^{{\rm i}Q\phi} &
-Q\theta\theta^{}_{\rm q}m^{}_{2}e^{{\rm i}Q\phi} \cr
-\theta^{}_{\rm q} m^{}_{2} e^{-{\rm i}Q\phi} &
m^{}_{2}+Q^{2}\theta^{2}m^{}_{3} & -Q\theta m^{}_{3} \cr
-Q\theta\theta^{}_{\rm q} m^{}_{2}e^{-{\rm i}Q\phi} & -Q\theta
m^{}_{3} & m^{}_{3} \end{matrix} \right) \; .
%   (25)
\end{eqnarray}
The hierarchical structure of $M^\prime_Q$ is therefore determined
by the hierarchy $|m^{}_1| \ll |m^{}_2| \ll m^{}_3$ and the smallness
of $\theta$ and $\theta^{}_{\rm q}$.

Corresponding to the numerical illustration of $M^{}_Q$ in Eqs.
(17)---(20), the results of $M^\prime_Q$ with the same inputs are
give below.

(a) $(m^{}_{1}, m^{}_2) = (-m^{}_u, +m^{}_c)$ for $Q=+2/3$ and
$(-m^{}_d, +m^{}_s)$ for $Q=-1/3$, leading to
\begin{eqnarray}
\begin{aligned}
M^{\prime}_{+2/3} \simeq \left(\begin{matrix} 3.337 &
-54.695e^{1.091{\rm i}} & -1.532e^{1.091{\rm i}} \cr
-54.695e^{-1.091{\rm i}} & 767.678 & -4798.559 \cr
-1.532e^{-1.091{\rm i}} & -4798.559 & 171965.605
\end{matrix}\right) {\rm MeV} \;,
%   (26)
\end{aligned}
\end{eqnarray}
and
\begin{eqnarray}
\begin{aligned}
M^{\prime}_{-1/3} \simeq \left(\begin{matrix} -0.317 &
-11.976e^{-0.545{\rm i}} & 0.168e^{-0.545{\rm i}} \cr
-11.976e^{0.545{\rm i}} & 55.047 & 39.272 \cr 0.168e^{0.545i} &
39.272 & 2859.450\end{matrix}\right){\rm MeV} \; ;
%   (27)
\end{aligned}
\end{eqnarray}

(b) $(m^{}_{1}, m^{}_2) = (+m^{}_u, +m^{}_c)$ for $Q=+2/3$ and
$(+m^{}_d, +m^{}_s)$ for $Q=-1/3$, leading to
\begin{eqnarray}
\begin{aligned}
M^{\prime}_{+2/3} \simeq \left(\begin{matrix} 6.077 &
-54.458e^{1.091{\rm i}} & -1.525e^{1.091{\rm i}} \cr
-54.458e^{-1.091{\rm i}} & 767.698 & -4798.559 \cr
-1.525e^{-1.091{\rm i}} & -4798.559 & 171965.605
\end{matrix}\right) {\rm MeV} \; ,
%   (28)
\end{aligned}
\end{eqnarray}
and
\begin{eqnarray}
\begin{aligned}
M^\prime_{-1/3} \simeq \left(\begin{matrix} 5.087 &
-10.847e^{-0.545{\rm i}} & 0.152e^{-0.545{\rm i}} \cr
-10.847e^{0.545{\rm i}} & 55.283 & 39.269 \cr 0.152e^{0.545i} &
39.269 & 2859.450
\end{matrix}\right){\rm MeV} \; .
%   (29)
\end{aligned}
\end{eqnarray}
One can see that the sign ambiguities of $m^{}_1$ and $m^{}_2$
mainly affect the magnitude of the $(1,1)$ element of $M^\prime_Q$.
The smallness of this matrix element is especially guaranteed if
$m^{}_1$ and $m^{}_2$ take the opposite signs, as numerically shown
in Eqs. (26) and (27).

In the hierarchy basis the language of texture ``zeros" has proved
to be very useful in establishing some experimentally testable
relations between the ratios of quark masses and the flavor mixing
angles \cite{F77,F78}. Those zeros dynamically mean that the
corresponding matrix elements are sufficiently suppressed as
compared with their neighboring counterparts, and this kind of
suppression may reasonably arise from an underlying flavor symmetry
\cite{FN}. In this sense Eqs. (26)---(29) motivate us to conjecture
the well-known four-zero textures of Hermitian quark mass matrices
\cite{Du} as the fairest extension of the original Fritzsch ansatz
which contains six texture zeros \cite{F78}:
\begin{eqnarray}
M^\prime_{Q} = \left(\begin{matrix} {\bf 0} & \diamondsuit^{}_Q &
{\bf 0} \cr \diamondsuit^*_Q & \heartsuit^{}_Q & \triangle^{}_Q \cr
{\bf 0} & \triangle^{*}_Q & \Box^{}_Q \cr\end{matrix}\right) \; ,
%     (30)
\end{eqnarray}
where the relevant symbols denote the nonzero matrix elements. In
fact, the pattern of $M^{}_Q$ with an approximate flavor democracy
obtained in Eq. (21) just leads us to the four-zero textures of
$M^\prime_Q$ in the hierarchy basis, if one takes $r \simeq 2 -
3\sqrt{2} \ Q\theta$ instead of $r \simeq 2$:
\begin{eqnarray}
M^\prime_Q \simeq \left(\begin{matrix} {\bf 0} &
-\theta^{}_{\rm q} m^{}_{2} e^{{\rm i}Q\phi} &
{\bf 0} \cr -\theta^{}_{\rm q}
m^{}_{2} e^{-{\rm i}Q\phi} & m^{}_{2}+Q^{2}\theta^{2}m^{}_{3} &
-Q\theta m^{}_{3} \cr {\bf 0} & -Q\theta m^{}_{3} & m^{}_{3}
\end{matrix} \right) \; ,
%   (31)
\end{eqnarray}
which can also be read off from Eq. (25) if similar approximations
are made. As pointed out in Refs. \cite{FX2003,XZ2015}, current
experimental data require that the (2,2) and (2,3) elements of
$M^\prime_{-1/3}$ be comparable in magnitude. In any case the
pattern of $M^{}_Q$ in Eq. (21) or the texture of $M^\prime_Q$ in
Eq. (31) can be very helpful for building a viable quark mass model.

\section{On the scale dependence}

In the above discussions we have restricted ourselves to the quark
mass matrices at the electroweak scale characterized by $\mu =
M^{}_Z$. Since the flavor democracy might be realized at a much
higher energy scale $M^{}_X$, where a kind of fundamental new
physics may occur, it makes sense to study the scale dependence of
$M^{}_Q$ by means of the one-loop renormalization-group equations
(RGEs) for the Yukawa coupling matrices and the CKM flavor mixing
matrix \cite{RGE1}. For the sake of simplicity, here we work in the
framework of the minimal supersymmetric standard model (MSSM) and
calculate the relevant RGEs by taking account of the strong
hierarchies of charged fermion masses and that of the CKM
parameters. The approximate analytical results turn out to be
\cite{RGE2}
\begin{eqnarray}
&& m^{}_{t}(M^{}_{Z}) \simeq m^{}_{t}(M^{}_{X})
\left(\zeta^{}_{\rm u} \xi^{6}_{t} \xi^{}_{b}\right) \;,
\nonumber \\
&& m^{}_{b}(M^{}_{Z}) \simeq m^{}_{b}(M^{}_{X}) \left(\zeta^{}_{\rm
d} \xi^{}_{t} \xi^{6}_{b} \xi^{}_{\tau}\right) \;; \hspace{0.8cm}
%     (32)
\end{eqnarray}
and
\begin{eqnarray}
\begin{aligned}
& \frac{ m^{}_{u} (M^{}_{X}) }{ m^{}_{t} (M^{}_{X}) } \simeq
\frac{ m^{}_{u} (M^{}_{Z}) }{ m^{}_{t} (M^{}_{Z}) }
\left(\xi^{3}_{t} \xi^{}_{b}\right) \;,
\\
& \frac{ m^{}_{c} (M^{}_{X}) }{ m^{}_{t} (M^{}_{X}) } \simeq
\frac{ m^{}_{c} (M^{}_{Z}) }{ m^{}_{t} (M^{}_{Z}) }
\left(\xi^{3}_{t} \xi^{}_{b}\right) \;,
\\
& \frac{ m^{}_{d} (M^{}_{X}) }{ m^{}_{b} (M^{}_{X}) } \simeq
\frac{ m^{}_{d} (M^{}_{Z}) }{ m^{}_{b} (M^{}_{Z}) }
\left(\xi^{}_{t} \xi^{3}_{b}\right) \;,
\\
& \frac{ m^{}_{s} (M^{}_{X}) }{ m^{}_{b} (M^{}_{X}) } \simeq
\frac{ m^{}_{s} (M^{}_{Z}) }{ m^{}_{b} (M^{}_{Z}) }
\left(\xi^{}_{t} \xi^{3}_{b}\right) \; ;
%   (33)
\end{aligned}
\end{eqnarray}
and
\begin{eqnarray}
\begin{aligned}
\theta^{}_{\rm u} (M^{}_{X}) \simeq \theta^{}_{\rm u} (M^{}_{Z}) \;,
~~~ \theta^{}_{\rm d} (M^{}_{X}) \simeq \theta^{}_{\rm d} (M^{}_{Z})
\;, ~~~ \theta (M^{}_{X}) \simeq \theta (M^{}_{Z}) \left(\xi^{}_{t}
\xi^{}_{b}\right) \;, ~~~ \phi (M^{}_{X}) \simeq \phi (M^{}_{Z}) \;
, ~~
%   (34)
\end{aligned}
\end{eqnarray}
where
\begin{eqnarray}
\begin{aligned}
&\zeta^{}_{\rm q} \equiv \exp \left[ \frac{1}{2}
\int^{{\rm ln}(M^{}_{X}/M^{}_{Z})}_{0} \sum^{3}_{i=1}
\frac{c^{\rm q}_{i} g^{2}_{i} (0)}{8\pi^{2} - b^{}_{i}g^{2}_{i} (0)
\chi} {\rm d} \chi \right] \;,
\\
&\xi^{}_{\alpha} \equiv \exp \left[ - \frac{1}{16\pi^2} \int^{{\rm
ln}(M^{}_{X}/M^{}_{Z})}_{0} f^{2}_{\alpha}(\chi) {\rm d} \chi
\right] \;
%   (35)
\end{aligned}
\end{eqnarray}
with $\rm q = u$ or $\rm d$, $\alpha = t$, $b$ or $\tau$, and $\chi
= {\rm ln}(\mu / M^{}_{Z})$. In Eq. (35) $c^{\rm q}_{i}$ and
$b^{}_{i}$ are the model-dependent coefficients whose values can be
found in Ref. \cite{RGE1}.

With the help of Eqs. (13) and (32)---(34), one can then express the
democratic quark mass matrices at $M^{}_{X}$ by using the quark
masses and flavor mixing parameters at $M^{}_{Z}$ and taking into
account their RGE evolution effects:
\begin{eqnarray}
\begin{aligned}
M^{}_{+2/3}(M^{}_{X}) & \simeq \frac{m^{}_{t}}
{3\zeta^{}_{\rm u} \xi^{6}_{t} \xi^{}_{b}} \left\{\left(
\begin{matrix} 1 & 1 & 1 \cr 1 & 1 & 1 \cr 1 & 1 & 1 \end{matrix}
\right)
\right. + \xi^{}_t \xi^{}_b \left[ \frac{1}{2} \xi^{2}_{t}
\frac{m^{}_{c}}{m^{}_{t}} \left( \begin{matrix} 1 & 1 & -2 \cr
1 & 1 & -2 \cr -2 & -2 & 4 \end{matrix} \right) + \frac{\sqrt{2}}{2}
\theta \left( \begin{matrix} 0 & 0 & 2 \cr
0 & 0 & 2 \cr 2 & 2 & 4 \end{matrix} \right) \right]
\\
&-\sqrt{3} \xi^{3}_{t} \xi^{}_{b} \theta^{}_{\rm u}
\frac{m^{}_{c}}{m^{}_{t}} \left[ \cos{ \left( + \frac{2}{3} \phi
\right)} \left( \begin{matrix} 1 & 0 & -1 \cr 0 & -1 & 1 \cr -1 & 1
& 0 \end{matrix} \right) + {\rm i} \sin{ \left(+ \frac{2}{3} \phi
\right)} \left(
\begin{matrix} 0 & 1 & -1 \cr -1 & 0 & 1 \cr 1 & -1 & 0 \end{matrix}
\right) \right]
\\
&-\frac{\sqrt{6}}{3} \xi^{4}_{t} \xi^{2}_{b} \theta \theta^{}_{\rm
u} \frac{m^{}_{c}}{m^{}_{t}} \left[\cos{ \left(+ \frac{2}{3} \phi
\right)} \left( \begin{matrix} 2 & 0 & 1 \cr 0& -2 & -1 \cr 1 & -1 &
0 \end{matrix} \right) - {\rm i} \sin{ \left(+ \frac{2}{3} \phi
\right)} \left(
\begin{matrix} 0 & -2 & -1 \cr 2 & 0 & 1 \cr 1 & -1 & 0 \end{matrix}
\right)\right]
\\
&\left.+\frac{3}{2} \xi^{3}_{t} \xi^{}_{b} \left(
\frac{m^{}_{u}}{m^{}_{t}} + \theta^{2}_{\rm u} \frac{m^{}_{c}}{m^{}_{t}}
\right)\left(\begin{matrix} 1 & -1 & 0 \cr -1 & 1 & 0 \cr 0 & 0 & 0
\end{matrix} \right)\right\} \;,
\end{aligned}
%   (36)
\end{eqnarray}
and
\begin{eqnarray}
\begin{aligned}
M^{}_{-1/3}(M^{}_{X}) & \simeq \frac{m^{}_{b}}
{3\zeta^{}_{\rm d} \xi^{}_{t} \xi^{6}_{b}\xi^{}_\tau} \left\{\left(
\begin{matrix} 1 & 1 & 1 \cr 1 & 1 & 1 \cr 1 & 1 & 1 \end{matrix} \right)
\right. + \xi^{}_t\xi^{}_b \left[ \frac{1}{2} \xi^{2}_{b}
\frac{m^{}_{s}}{m^{}_{b}} \left( \begin{matrix} 1 & 1 & -2 \cr 1 & 1
& -2 \cr -2 & -2 & 4 \end{matrix} \right) - \frac{\sqrt{2}}{4}
\theta \left( \begin{matrix} 0 & 0 & 2 \cr 0 & 0 & 2 \cr 2 & 2 & 4
\end{matrix} \right)\right]
\\
&-\sqrt{3} \xi^{}_{t} \xi^{3}_{b} \theta^{}_{\rm d}
\frac{m^{}_{s}}{m^{}_{b}} \left[ \cos{ \left( - \frac{1}{3} \phi
\right)} \left( \begin{matrix} 1 & 0 & -1 \cr 0 & -1 & 1 \cr -1 & 1
& 0 \end{matrix} \right) + {\rm i} \sin{ \left(- \frac{1}{3} \phi
\right)} \left( \begin{matrix} 0 & 1 & -1 \cr -1 & 0 & 1 \cr 1 & -1
& 0 \end{matrix} \right) \right]
\\
&+ \frac{\sqrt{6}}{6} \xi^{2}_{t} \xi^{4}_{b} \theta \theta^{}_{\rm
d} \frac{m^{}_{s}}{m^{}_{b}} \left[\cos{ \left(- \frac{1}{3} \phi
\right)} \left( \begin{matrix} 2 & 0 & 1 \cr 0& -2 & -1 \cr 1 & -1 &
0 \end{matrix} \right) - {\rm i} \sin{ \left(- \frac{1}{3} \phi
\right)} \left( \begin{matrix} 0 & -2 & -1 \cr 2 & 0 & 1 \cr 1 & -1
& 0 \end{matrix} \right) \right]
\\
&\left. + \frac{3}{2} \xi^{}_{t} \xi^{3}_{b} \left(
\frac{m^{}_{d}}{m^{}_{b}} + \theta^{2}_{\rm d}
\frac{m^{}_{s}}{m^{}_{b}} \right)\left(\begin{matrix} 1 & -1 & 0 \cr
-1 & 1 & 0 \cr 0 & 0 & 0 \end{matrix} \right)\right\} \;,
\end{aligned}
%   (37)
\end{eqnarray}
where $r^{}_{Q} \simeq 2$ has been taken. Typically taking
$M^{}_{X}=10^{16} ~{\rm GeV}$, $M^{}_{Z}=91.187 ~{\rm GeV}$ and
$\tan{\beta^{}_{\rm MSSM}}=10$ for illustration, we numerically
obtain $\zeta^{}_{\rm u} \simeq 3.47$, $\zeta^{}_{\rm d} \simeq
3.38$, $\xi^{}_{t} \simeq 0.854$, $\xi^{}_{b} \simeq 0.997$ and
$\xi^{}_{\tau} \simeq 0.998$ from the one-loop RGEs \cite{RGE2}. In
this case the expressions of $M^{}_{+2/3}$ and $M^{}_{-1/3}$ at
$M^{}_X$ turn out to be
\begin{eqnarray}
\begin{aligned}
M^{}_{+2/3}(M^{}_{X}) & \simeq 0.75 \cdot \frac{1}{3} m^{}_{t}
\left\{\left(
\begin{matrix} 1 & 1 & 1 \cr 1 & 1 & 1 \cr 1 & 1 & 1 \end{matrix}
\right)\right.
+ 0.85 \left[ 0.73 \cdot \frac{1}{2} \frac{m^{}_{c}}{m^{}_{t}}
\left( \begin{matrix} 1 & 1 & -2 \cr 1 & 1 & -2 \cr -2 & -2 & 4
\end{matrix} \right) + \frac{ \sqrt{2}}{2} \theta \left(
\begin{matrix} 0 & 0 & 2 \cr 0 & 0 & 2 \cr 2 & 2 & 4
\end{matrix} \right) \right]
\\
&- 0.62 \cdot \sqrt{3} \theta^{}_{\rm u} \frac{m^{}_{c}}{m^{}_{t}}
\left[ \cos{ \left( + \frac{2}{3} \phi \right)} \left(
\begin{matrix} 1 & 0 & -1 \cr 0 & -1 & 1 \cr -1 & 1 & 0 \end{matrix}
\right) + {\rm i} \sin{ \left(+ \frac{2}{3} \phi \right)} \left(
\begin{matrix} 0 & 1 & -1 \cr -1 & 0 & 1 \cr 1 & -1 & 0
\end{matrix} \right) \right]
\\
&- 0.53 \cdot \frac{\sqrt{6}}{3} \theta \theta^{}_{\rm u}
\frac{m^{}_{c}}{m^{}_{t}} \left[\cos{ \left(+ \frac{2}{3} \phi
\right)} \left( \begin{matrix} 2 & 0 & 1 \cr 0 & -2 & -1 \cr 1 & -1
& 0 \end{matrix} \right) - {\rm i} \sin{ \left(+\frac{2}{3}
\phi\right)} \left( \begin{matrix} 0 & -2 & -1 \cr 2 & 0 & 1 \cr 1 &
-1 & 0
\end{matrix}\right)\right]
\\
&\left.+ 0.62 \cdot \frac{3}{2} \left( \frac{m^{}_{u}}{m^{}_{t}} +
\theta^{2}_{\rm u} \frac{m^{}_{c}}{m^{}_{t}}
\right)\left(\begin{matrix} 1 & -1 & 0\cr -1 & 1 & 0 \cr 0 & 0 & 0
\end{matrix} \right)\right\} \;,
\end{aligned}
%   (38)
\end{eqnarray}
and
\begin{eqnarray}
\begin{aligned}
M^{}_{-1/3}(M^{}_{X}) & \simeq 0.35 \cdot \frac{1}{3} m^{}_{b}
\left\{\left(
\begin{matrix} 1 & 1 & 1 \cr 1 & 1 & 1 \cr 1 & 1 & 1 \end{matrix}
\right)
\right. + 0.85 \left[ 1.00 \cdot \frac{1}{2}
\frac{m^{}_{s}}{m^{}_{b}} \left(
\begin{matrix} 1 & 1 & -2 \cr 1 & 1 & -2 \cr -2 & -2 & 4 \end{matrix}
\right)
- \frac{\sqrt{2}}{4} \theta \left( \begin{matrix} 0 & 0 & 2 \cr 0 & 0 & 2
\cr 2 & 2 & 4 \end{matrix} \right) \right]
\\
&- 0.85 \cdot \sqrt{3} \theta^{}_{\rm d} \frac{m^{}_{s}}{m^{}_{b}}
\left[ \cos{\left( - \frac{1}{3} \phi \right)} \left( \begin{matrix}
1 & 0 & -1 \cr 0 & -1 & 1 \cr -1 & 1 & 0 \end{matrix} \right) + {\rm
i} \sin{ \left(- \frac{1}{3} \phi \right)} \left( \begin{matrix} 0 &
1 & -1 \cr -1 & 0 & 1 \cr 1 & -1 & 0 \end{matrix} \right) \right]
\\
&+ 0.72 \cdot \frac{\sqrt{6}}{6} \theta \theta^{}_{\rm d}
\frac{m^{}_{s}}{m^{}_{b}} \left[\cos{ \left(- \frac{1}{3} \phi
\right)} \left( \begin{matrix} 2 & 0 & 1 \cr 0 & -2 & -1 \cr 1 & -1
& 0 \end{matrix} \right) - {\rm i} \sin{ \left(- \frac{1}{3} \phi
\right)} \left( \begin{matrix} 0 & -2 & -1 \cr 2 & 0 & 1 \cr 1 & -1
& 0
\end{matrix} \right) \right]
\\
&\left. + 0.85 \cdot \frac{3}{2} \left( \frac{m^{}_{d}}{m^{}_{b}} +
\theta^{2}_{\rm d} \frac{m^{}_{s}}{m^{}_{b}}
\right)\left(\begin{matrix} 1 & -1 & 0\cr -1 & 1 & 0 \cr 0 & 0 & 0
\end{matrix} \right)\right\}  \;,
\end{aligned}
%   (39)
\end{eqnarray}
from which one can clearly see the RGE-induced corrections to the
relevant terms in each quark sector. Hence such quantum effects
should not be ignored when building a specific quark mass model
based on the flavor democracy at $M^{}_X$ and confronting its
predictions with the experimental data at $M^{}_Z$.

At this point it is worth mentioning that the approximate four-zero
textures of $M^{}_{+2/3}$ and $M^{}_{-1/3}$ in the hierarchy basis
are essentially stable against the RGE running effects. Here the
stability of the texture zeros means that the (1,1), (1,3) and (3,1)
elements of each quark mass matrix at $M^{}_X$ remain strongly
suppressed in magnitude as compared with their neighboring
counterparts, and thus it is a reasonable approximation to take them
to be vanishing at any energy scale between $M^{}_Z$ and $M^{}_X$
from a phenomenological point of view \cite{XZ2015}. Such an
observation makes sense because the four-zero textures of Hermitian
quark mass matrices or their variations are especially favored by
current experimental data and deserve some special attention in the
model-building exercises.

\section{Summary}

It has been known for quite a long time that the democracy of quark
flavors is one of the well-motivated flavor symmetries for building
a viable quark mass model, but how to break this symmetry and to
what extent to break it are highly nontrivial. To minimize the
number of free parameters, in this work we have assumed the
structural parallelism between $Q=+2/3$ and $Q=-1/3$ quark sectors,
and proposed a novel way to reconstruct the texture of flavor
democracy breaking and evaluate its strength in each sector with the
help of the Fritzsch-Xing parametrization of the CKM flavor mixing
matrix. Some phenomenological implications of such
flavor-democratized quark mass matrices, in particular their
variations with possible texture zeros in the hierarchy basis and
their RGE evolution from the electroweak scale to a superhigh-energy
scale, have also been discussed. We hope that this kind of study
will be useful to more deeply explore the underlying correlation
between the quark flavor structures and the observed quark mass
spectrum and flavor mixing pattern.

\vspace{0.3cm}

This research work was supported
in part by the National Natural Science Foundation of China under
grant No. 11375207 and the National Basic Research Program of
China under grant No. 2013CB834300.

\end{document}